\documentclass[preprint,prd,tightenlines,nofootinbib,floatfix]{revtex4}
\usepackage{graphicx}

\def \MSbar {\vbox{\hrule\kern 1pt\hbox{\rm MS}}}

\begin{document}

\begin{center}
{\Large   On the study of the QCD string model and its application to meson spectroscopy
}\\[2ex]
{Y. H. Yuan
\footnote{E-mail:henry@physics.wisc.edu }
}

Department of Physics,University of Wisconsin,Madison, WI 53706

\end{center}

\begin{abstract}

The meson spectroscopy is studied in the framework of the QCD string model. The mesons are composed of spinless quarks which move relativistically. We show that the QCD string model reproduces the straight line Regge trajectories and exibit the correct string slopes for both light-light and heavy-light mesons. We also present the exact numerical solution and the analytic approximation solution of the QCD string equations for the light-light mesons under the deep radial limit. We demonstrate that the QCD string model describes the universal relations for the light-light and heavy-light meson spectroscopies correctly from the comparison of the exact numerical solutions of the QCD string model for both light-light and heavy-light mesons.
\end{abstract}

\maketitle

\section{Introduction}

The mass spectrum of a hadron is a fundamental problem in hadronic physics. The QCD motivated potential model with scalar confinement in the nonrelativistic limit has been employed to deal with this problem. However the scalar potential confinement behaves badly\cite{co} in the limit of small quark mass and diverges from QCD once we consider the relativistic corrections. It was found that the spin independent relativistic corrections using the low velocity Wilson loop formalism differed from those of scalar confinement\cite{Milan}. The flux tube model which plays an important role in describing the relativistic meson states extends the nonrelativistic constituent quark model to include the gluonic degree of freedom. The review of the flux tube model may be found in Swanson's paper\cite{sw}. The QCD string or the relativistic flux tube(RFT) model which is the simplest generalization of the potential model overcomes these shortcomings of the scalar confinement. Its fundamental assumption is that the QCD dynamical ground state for large quark separation consists of a rigid straight tube-like color flux structure connecting the quarks and for massless quarks it reduces to Nambu string.\cite{na} This model is consistent with both spin-dependent and spin-independent QCD expectations while the scalar confinement potential model\cite{bu} is not consistent with spin-independent QCD relativistic correction. It includes the angular momentum and the rotational energy of the interacting field while only the field energy is considered in the potential model. Even though with this simple physical picture the RFT model coincides with the QCD relativistic correction derived from the rigorous Wilson loop formalism. Next we review the straight QCD string model and give the QCD string equations of mesons. Consider a light-light(LL) meson made up of two spinless quarks with equal mass $m_l$. The quarks may rotate around their center of mass and vibrate in the radial direction. We obtain the following equations of the QCD string\cite{lm} for the LL mesons,

\begin{eqnarray} 
\frac{2 L}{r}=2W_r \gamma_\bot v_\bot +2arf(v_\bot)
 \label{l}\\ 
E=2W_r \gamma_\bot +arS(v_\bot)+V(r)
\label{m} 
\end{eqnarray}

where $ l$ and $E $ are angular momentum and energy of the meson respectively, V(r) represents the short range potential, a is the confinement constant and r is the interquark separation. We also define the functions:

\begin{eqnarray} 
S(v_\bot) = \frac{arcsin \,v_\bot}{v_\bot}
\label{eq.3}\\
f(v_\bot)={1\over 4 v_\bot} [\,S(v_\bot) -\sqrt{1-v_\bot ^2}\,]
 \label{eq.4}\\
W_r =\sqrt{p_r ^2 +m^2}
\label{eq.5}\\
\gamma_\bot^2 = \frac{1}{1- v_\bot^2}
\label{eq.6}
\end{eqnarray}

Similarly, for a heavy-light meson(HL) made up of one heavy quark with mass M and one light quark with mass $ m_l$. They satisfy the conditions as $M\gg m_l\rightarrow 0$  and $ M\gg ar  $ , we obtain the equations of QCD string for HL mesons,

\begin{eqnarray} 
\frac Lr=W_r \gamma_\bot v_\bot +2arf(v_\bot)
 \label{e}\\ 
E=W_r \gamma_\bot +arS(v_\bot)+V(r)
\label{q} 
\end{eqnarray}

The paper is organized as follows. In section II we solve the RFT equations under two limits for the LL mesons. We demonstrate an analytical approximation solution under the deep radial limit in section III. The comparison of solutions for LL mesons with those for HL mesons is discussed in section IV. Special attentions are payed to the string slopes since they are significant criterions to tell whether a QCD motivated model is reasonable or not. Finally in section V our conclusions are summarized.

\section{Solutions under two limits}

\subsection{Circular Motion limit}

We now consider the circular motion limit for the LL mesons in which the two light spinless quarks moving circularly around their center of mass. So this will reduce the $p_r \rightarrow 0$ and $ v_\bot \rightarrow 1$. This reduction yields 

\begin{eqnarray} 
S(v_\bot)=\frac \pi2
\label{eq.8}\\
f(v_\bot)=\frac \pi8
 \label{eq.9}
\end{eqnarray}

Pluging functions $S(v_\bot), f(v_\bot)$ back into the general LL RFT equations, we obtain

\begin{eqnarray} 
\frac{2 L}{r}=\frac{\pi a r}{4}
 \label{1}\\ 
E=\frac{\pi a r}{2}
\label{2} 
\end{eqnarray}

Therefore we find the $`Nambu'$ slope for the LL meson is,

\begin{equation} 
{\alpha}^{'}_{LL} = \frac {L}{E^2}=\frac {1}{2\pi a}
\end{equation} 

We note that it is equal to half of that for HL meson. We will discuss this slope again in section \ref{comp} where the results can be seen clearly in the corresponding graphs no matter whether there is color Coulomb potential or not.

\subsection{The deep radial limit solutions for LL mesons}

Under the deep radial limit, considering the LL mesons and assuming again the light quark mass is zero we want to explore the effect of the short range color Coulomb potential. That the QCD string reduces to a time component vector potential was shown under the deep radial limit in which $L\ll E^2$.\cite{ao} It is reasonable to apply the semi-classical quantization condition for a spherically symmetric system to quantize the QCD string,

\begin{equation} 
(n+ \frac 12)\pi = \int_{r_-}^{r_+} dr p_r = I
\label{aa}
\end{equation} 

where $r_+, r_-$ represent the radial distances at the turning points of the motion. Assume the potential takes the simple form  $V(r)=-\frac kr$\cite{Henriques}. we obtain under the deep radial limit,

\begin{eqnarray} 
4p_r^2+ 4\frac{ L^2}{r^2}=(E-ar+\frac{k}{r})^2, 
\label{jj}   \\ 
p_r=\frac{1}{2r}\sqrt{(Er-ar^2-2L+k)(Er-ar^2+2L+k)}.
\label{ii}  
\end{eqnarray}

We now define the dimensionless variables,

\begin{eqnarray}
x=\sqrt{a}r         \label{xx}  \\
e=\frac{E} {2\sqrt{a} } \label{ww} \\
P_r=\frac{p_r}{\sqrt{a}} \label{vv} 
\end{eqnarray}

Then we obtain using the new dimensionless variables

\begin{equation}
P_r = \frac{p_r}{\sqrt{a}}             \nonumber  \\ 
 = \frac{\sqrt{(x^2-2ex+2L-k)(x^2-2ex-2L-k)}}{2x}  \nonumber  \\
 = \frac{\sqrt{Q_1Q_2}}{2x}
\end{equation} 

where 
\begin{eqnarray}
Q_1=x^2-2ex+2L-k, 
\label{uu}  \\
Q_2=x^2-2ex-2L-k.
\label{tt}  
\end{eqnarray}

The turning points are as follows:

\begin{eqnarray}
x_1=e-\sqrt{e^2+2L+k},\label{ss} \\
x_2=e-\sqrt{e^2-2L+k},  \label{rr}\\
x_3=e+\sqrt{e^2-2L+k}, \label{qq} \\ 
x_4=e+\sqrt{e^2+2L+k}, \label{pp} \\
\end{eqnarray}
 
Now we are ready to evaluate the integral

\begin{equation}
I \equiv \int^{r_+}_{r_-} dr p_r =\int^{x_3}_{x_2} dx \frac{\sqrt{Q_1Q_2}}{2x}
\label{iii}
\end{equation}
Assume,

\begin{equation}
x=uy+v
\end{equation} 

therefore,

\begin{equation} 
Q_1=(uy+v-x_2)(uy+v-x_3)   \nonumber \\
=u^2y^2+uy(2v-x_2-x_3)+(v-x_2)(v-x_3 ) \nonumber \\
=u^2y^2-(e^2-2L+k)
\end{equation} \\

where we let 
\begin{equation}
 v =\frac{x_2+x_3}{2}.
\end{equation}

Similarly, 

\begin{equation} 
Q_2=(uy+v-x_1)(uy+v-x_4) \\
=u^2y^2+uy(2v-x_1-x_4)+(v-x_1)(v-x_4 ) \\
=u^2y^2-(e^2+2L+k)
\end{equation} \\

Define

\begin{equation}
u^2=e^2-2L+k
\end{equation}

then,

\begin{equation} 
Q_1Q_2 = u^2(1-y^2)(1-my^2)(e^2+2L+k)
\end{equation} \\

where 
\begin{equation} 
m = \frac{ e^2-2L+k}{e^2+2L+k}
\end{equation} 

Due to the following relation,

\begin{eqnarray} 
x_2-e=-u  \label{oo} \\
x_3-e=u   \label{nn}
\end{eqnarray}

plug them into Eq.(\ref{iii})

\begin{eqnarray} 
I &=& \int^{x_3}_{x_2} dx \frac{\sqrt{Q_1Q_2}}{2x}  \nonumber  \\
&=& \frac{u^2}{2}\int^{1}_{-1} \frac{\displaystyle dy}{\displaystyle uy+e}\sqrt{(1-y^2)(1-my)(e^2+2L+k)} \nonumber  \\
&=& \frac{u^2\sqrt{e^2+2L+k}}{2}\Big{[}\int^{0}_{-1}\frac{\displaystyle dy}{\displaystyle uy+e}\sqrt{(1-y^2)(1-my^2)}+\int^{1}_{0}\frac{\displaystyle dy}{\displaystyle uy+e}\sqrt{(1-y^2)(1-my^2)}\,\Big{]} \nonumber  \\
&=& \frac{u^2\sqrt{e^2+2L+k}}{2}\int^{1}_{0} dy\Big{[}\sqrt{(1-y^2)(1-my^2)}(\frac{\displaystyle 1}{\displaystyle -uy+e}+\frac{\displaystyle 1}{\displaystyle uy+e})\,\Big{]} 
\label{mm}
\end{eqnarray}

Therefore we get the TCV integral for the LL mesons under the deep radial limit,

\begin{eqnarray} 
I &=& \frac{(e^2-2L+k) \sqrt{e^2+2L+k}}{e} \int^{1}_{0} dy \frac {\sqrt{(1-y^2)(1-\frac{\displaystyle e^2-2L+k}{\displaystyle e^2+2L+k}y^2)}}{1-\frac{\displaystyle e^2-2L+k}{\displaystyle e^2}y^2}   \nonumber  \\
&=& \frac{(e^2-2L+k) \sqrt{e^2+2L+k}}{e} J
\end{eqnarray} 

where 
 
\begin{equation}
J = \int^1_0 dy \frac {\sqrt{(1-y^2)(1-\frac{\displaystyle e^2-2L+k}{\displaystyle e^2+2L+k}y^2)}}{1-\frac{\displaystyle e^2-2L+k}{\displaystyle e^2}y^2}. 
\label{jjj}
\end{equation}

Using the semi-classical quantization condition for a spherically symmetric system to quantize the QCD string, we obtain,

\begin{equation} 
I =\frac{(e^2-2L+k) \sqrt{(e^2+2L+k)}}{e}\int^1_0 dy \frac {\sqrt{(1-y^2)(1-\frac{\displaystyle e^2-2L+k}{\displaystyle e^2+2L+k}y^2)}}{1-\frac{\displaystyle e^2-2L+k}{\displaystyle e^2}y^2}=(n+\frac12)\pi
\label{bb}
\end{equation} 

When evaluating Eq.(\ref{bb}) we need replace the classical angular momentum with Langer correction to take into account the centrifugal singularity, namely $L \to l+\frac12$. Moreover, this integral $I$ may be represented with complete standard elliptic integrals,

\begin{equation} 
I=\frac{(e^2-2L+k) \sqrt{(e^2+2L+k)}}{e} [(1-\frac{m}{h})(1-\frac1h){\Pi}(h,m) + \frac{m}{h}(1-\frac1h)F(m)+\frac1h E(m)]
\end{equation} 
where 

\begin{equation} 
h = \frac{e^2-2L+k}{e^2}. 
\end{equation} 

\section{Analytic approximation solution} \label{ana}

Next we derive an analytical approximation solution of QCD string equations for LL mesons under the deep radial limit $L\ll E^2$. After some tedious calculations (Shown in Appendix) we obtain an analytical approximation of the spectroscopic relation for LL mesons,

\begin{equation} 
L + 2 n+ \frac 32 =\frac{E^2}{2\pi a} 
\label{dd}
\end{equation} \\

We now compare this approximation result with the exact numerical solution Eq.(\ref{bb}). We draw the 6 figures for every other $n$ starting from 0 to 10. 
From these figures(from FIG.1 to FIG.6) we easily identify that the approximations are more accurate along with the increasement of n since the deep radial limit is better satisfied.

\begin{figure}[ht]
\includegraphics[width=\columnwidth,angle=0]{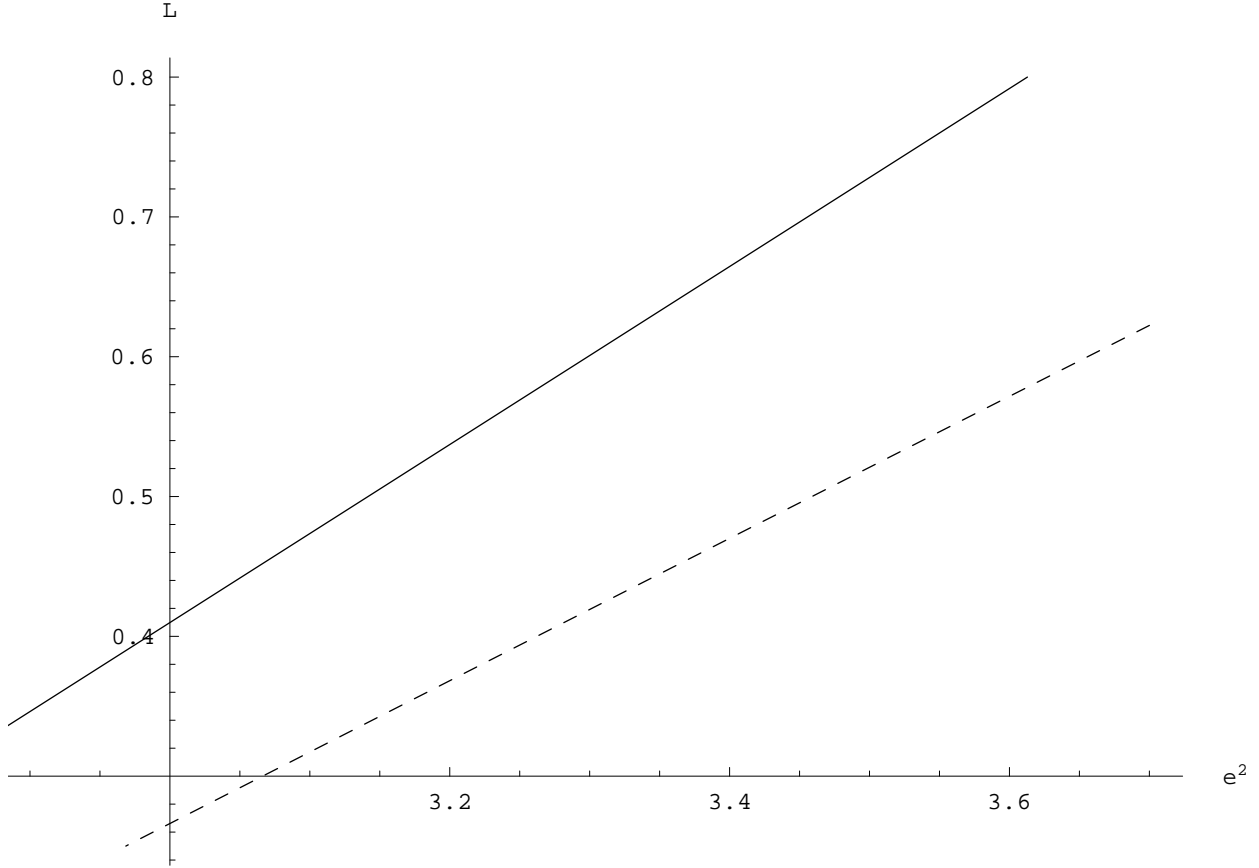}
\caption{ground state n=0 \label{fig:1}}
\end{figure}

\begin{figure}[ht]
\includegraphics[width=\columnwidth,angle=0]{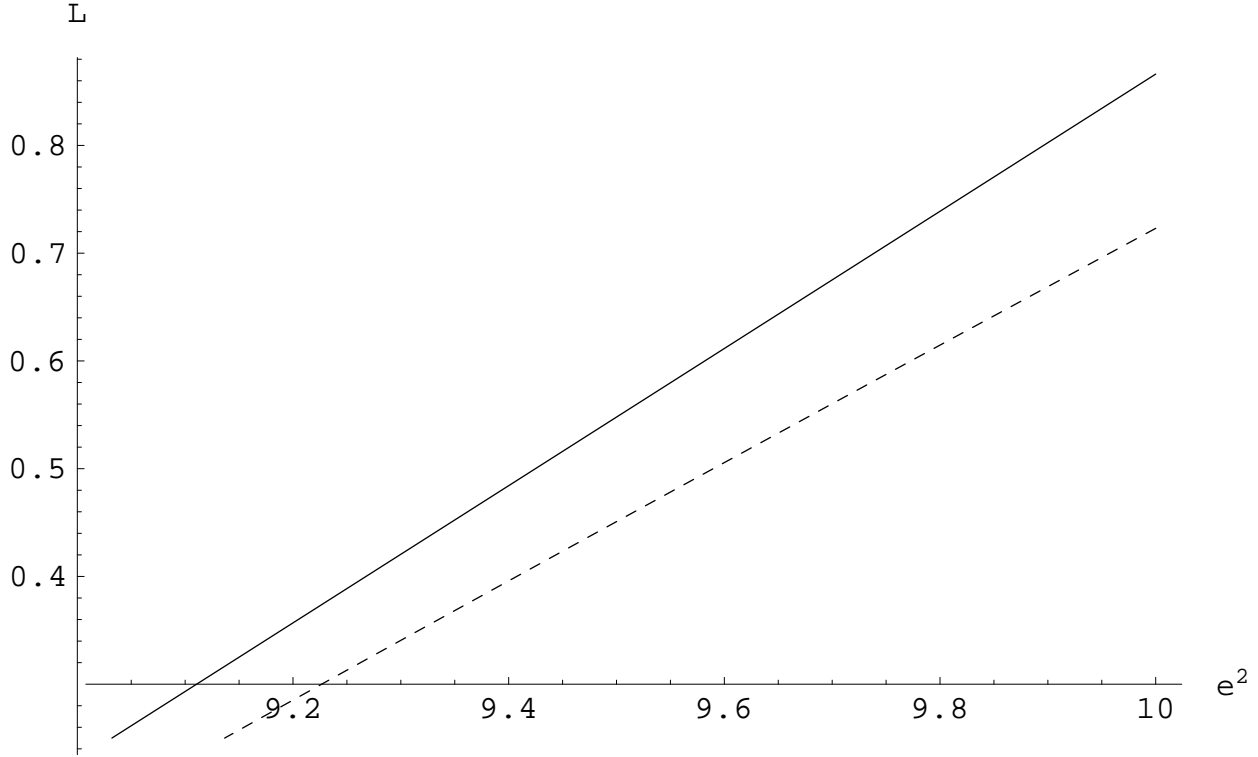}
\caption{The 2nd excited state \label{fig:2}}
\end{figure}

\begin{figure}[ht]
\includegraphics[width=\columnwidth,angle=0]{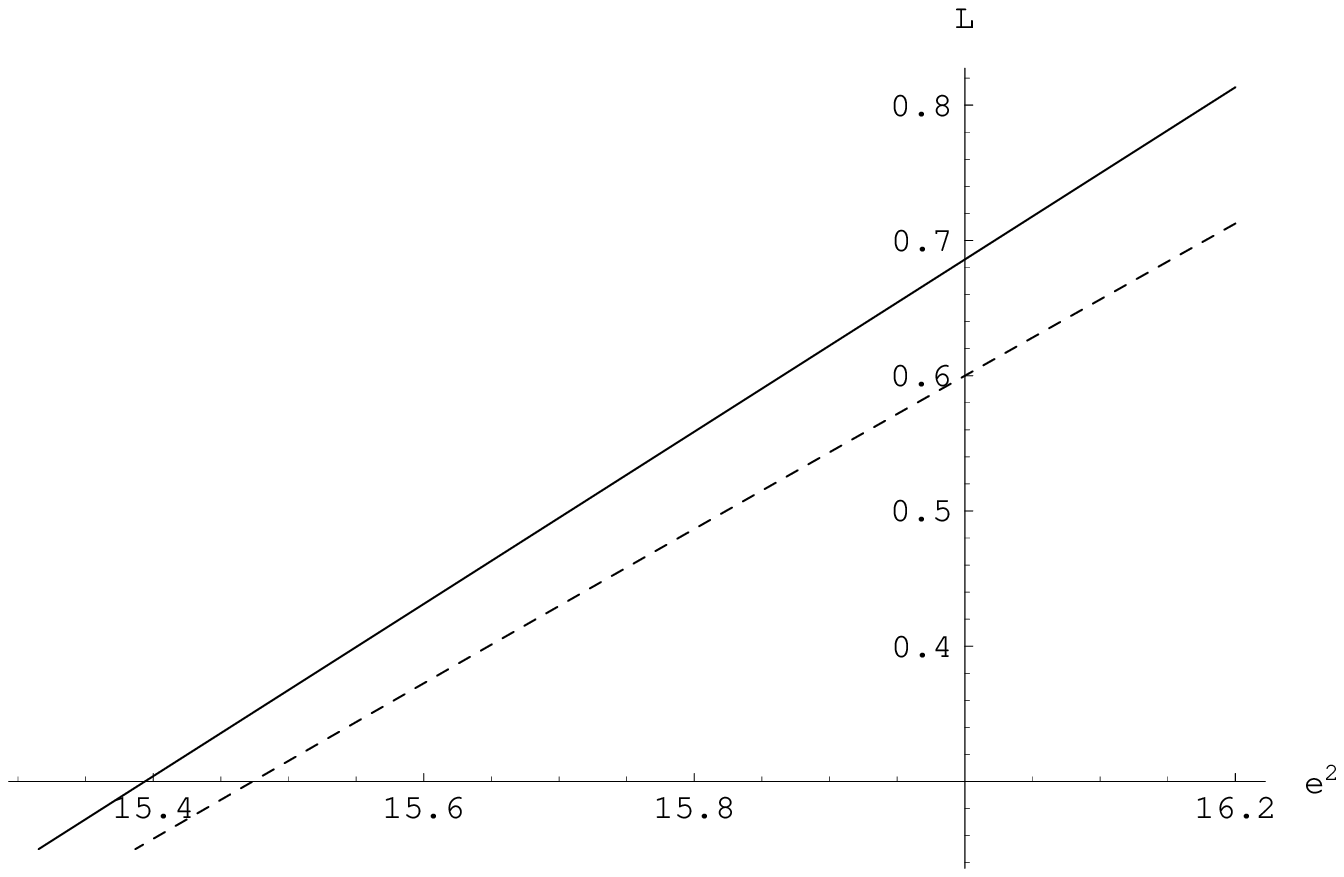}
\caption{The 4th excited state \label{fig:3}}
\end{figure}

\begin{figure}[ht]
\includegraphics[width=\columnwidth,angle=0]{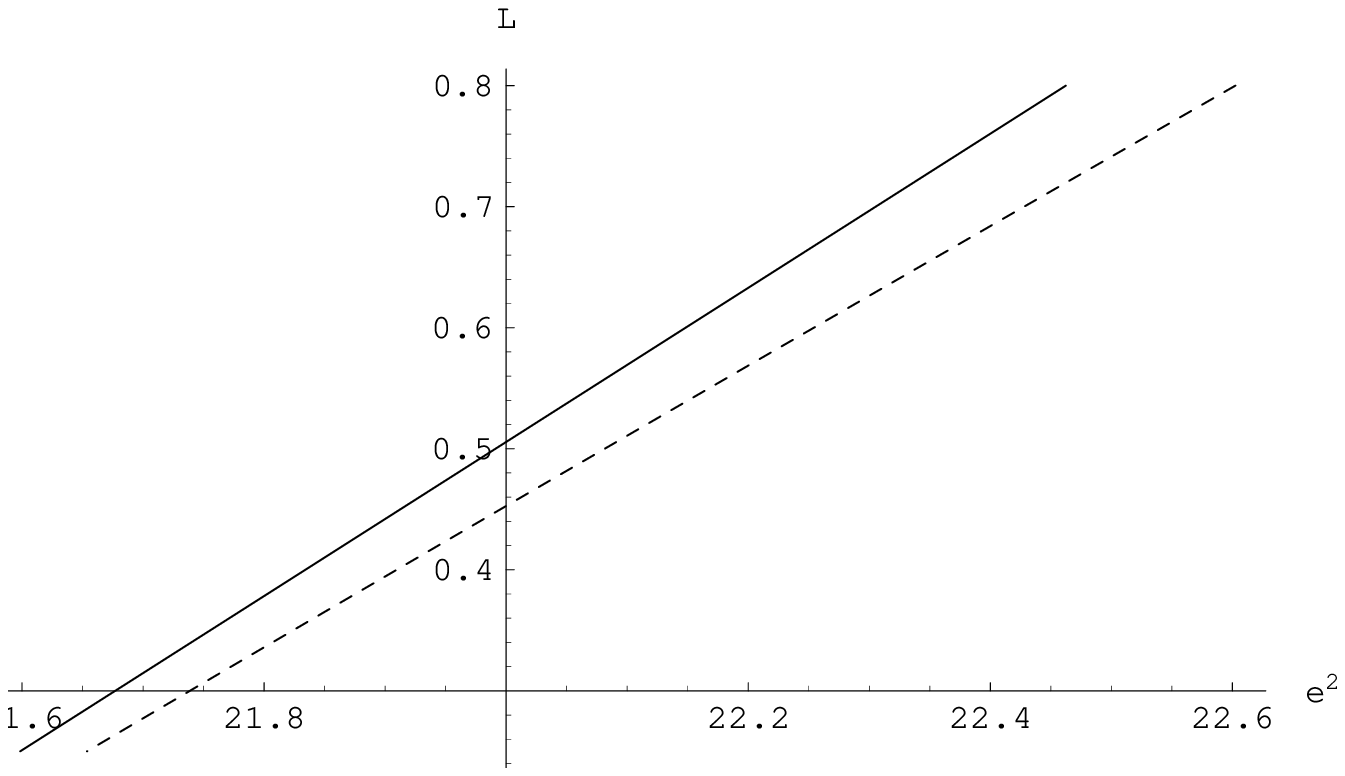}
\caption{The 6th excited state \label{fig:4}}
\end{figure}

\begin{figure}[ht]
\includegraphics[width=\columnwidth,angle=0]{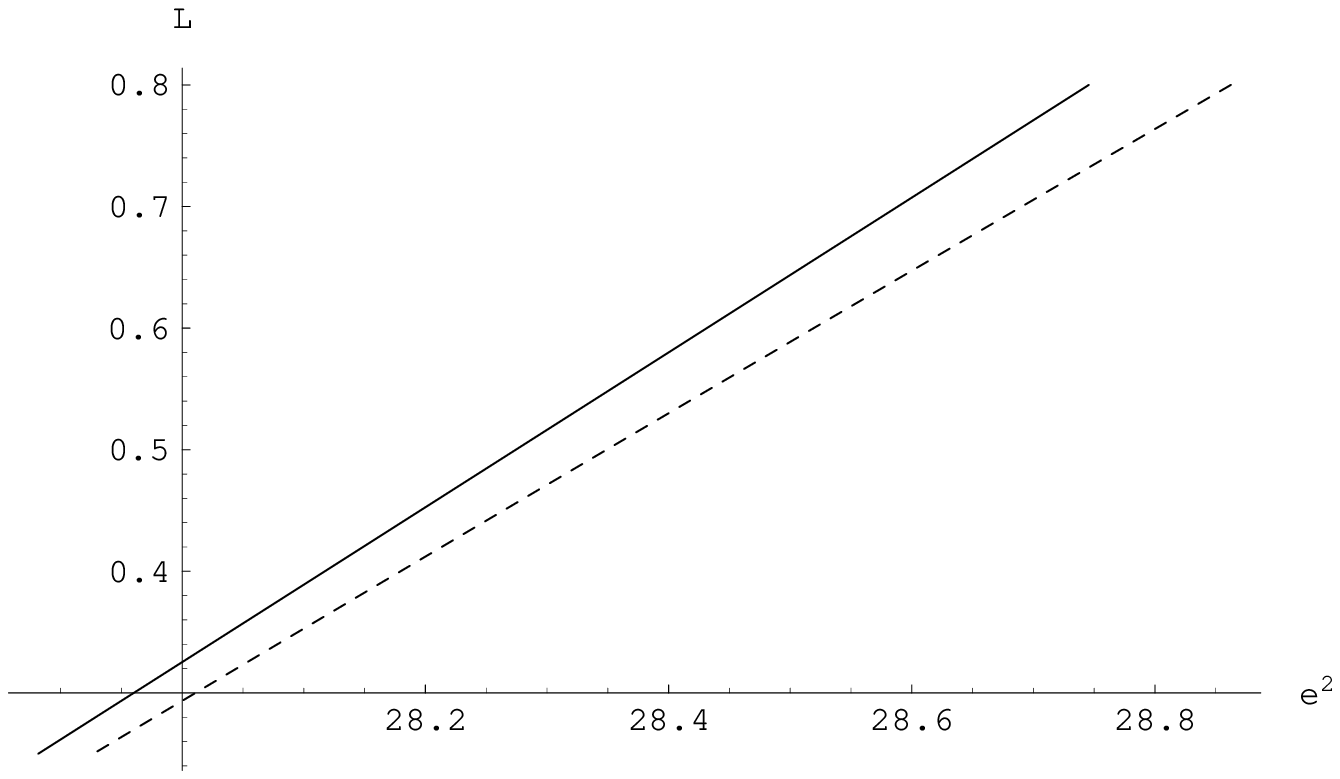}
\caption{The 8th excited state \label{fig:5}}
\end{figure}

\begin{figure}[ht]
\includegraphics[width=\columnwidth,angle=0]{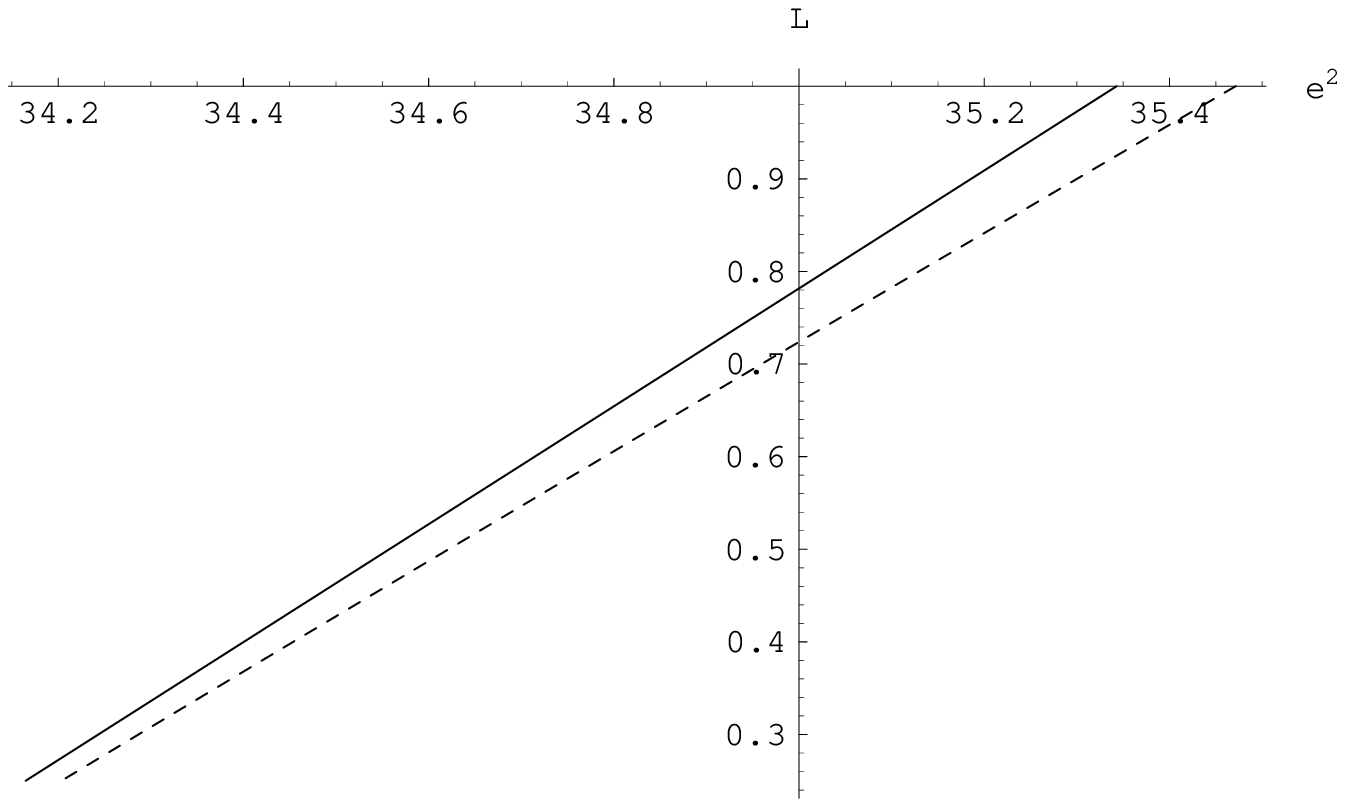}
\caption{The 10th excited state \label{fig:6}}
\end{figure}

\section{comparison of solutions for LL mesons with those for HL mesons under the deep radial limit} \label{comp}

\begin{figure}[ht]
\includegraphics[width=\columnwidth,angle=0]{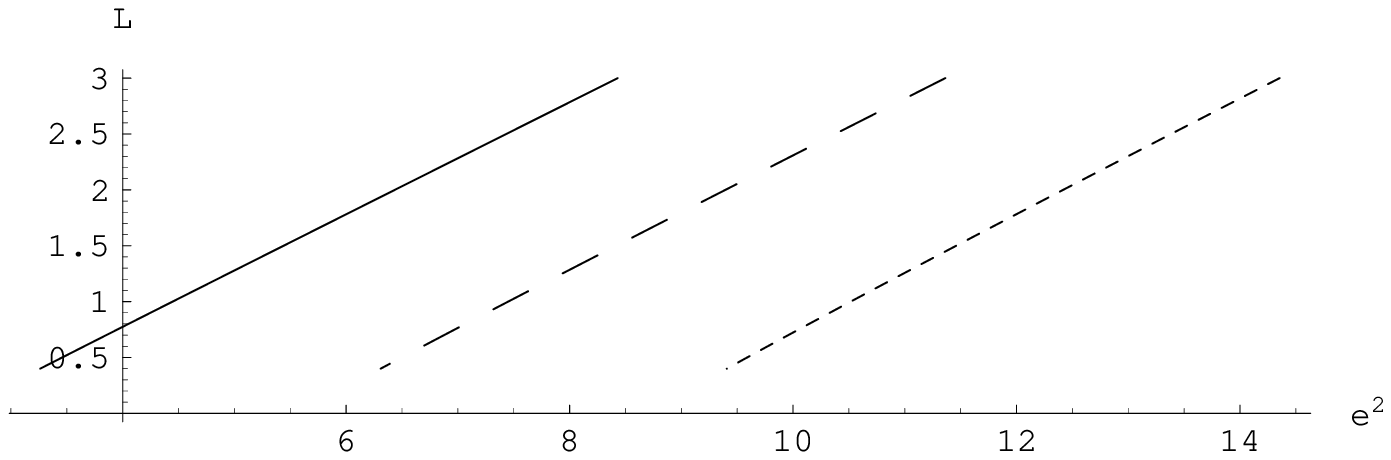}
\caption{The middle dashed line is for the first excited state while the right one is for the 2nd excited state. They compare with the ground state graph represented by the solid line. \label{17}}
\end{figure}

Next we take advantage of numerical methods to draw the graphs for LL meson integral. At first we turn off the color Coulomb potential. For the ground state, the 1st and 2nd excited states, the relation of angular momentum and energy is shown respectively in FIG.\ref{17}

\begin{figure}[ht]
\includegraphics[width=\columnwidth,angle=0]{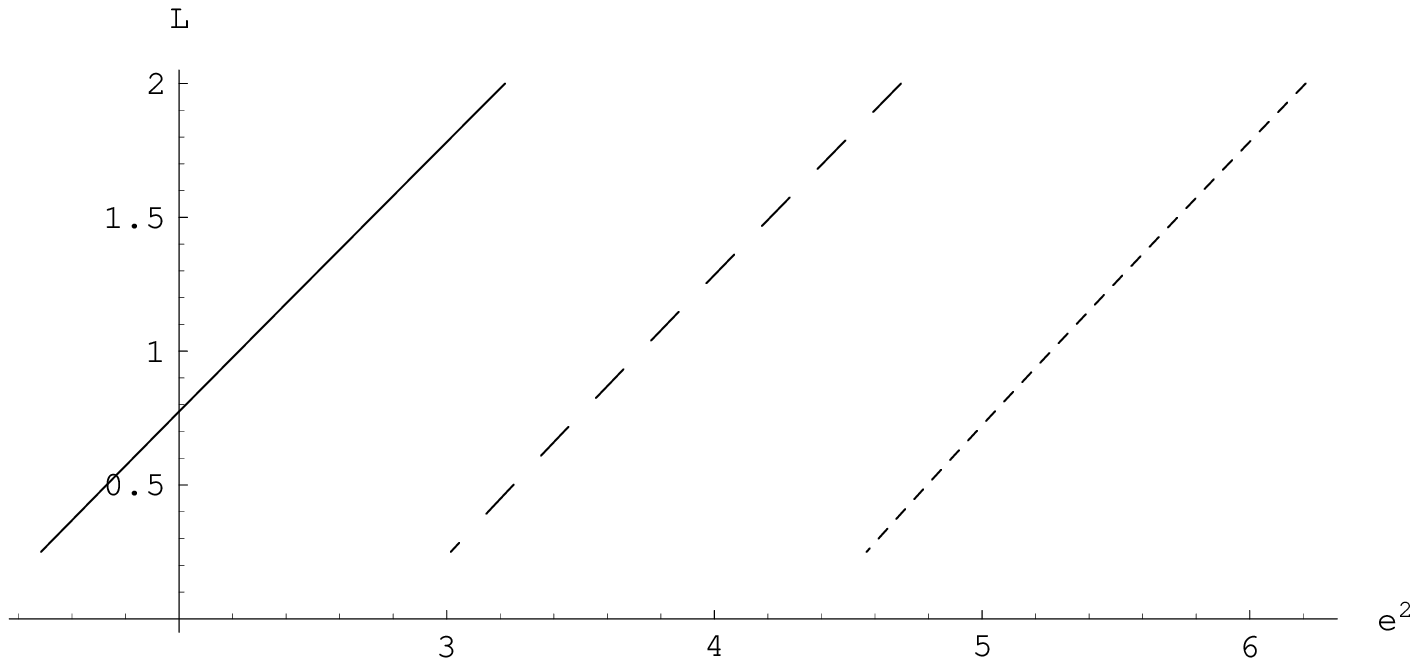}
\caption{The middle dashed line is for the first excited state, the right one is for the 2nd excited state while the solid line is for the ground state. \label{19}}
\end{figure}

From the FIG.\ref{17} we easily notice that the slopes of the curve for LL meson are half of those for HL meson which are shown in FIG.\ref{19}. The RFT mode reproduces the straight line Regge trajectories and exibit the correct string slopes for both LL and HL mesons.

\begin{figure}[ht]
\includegraphics[width=\columnwidth,angle=0]{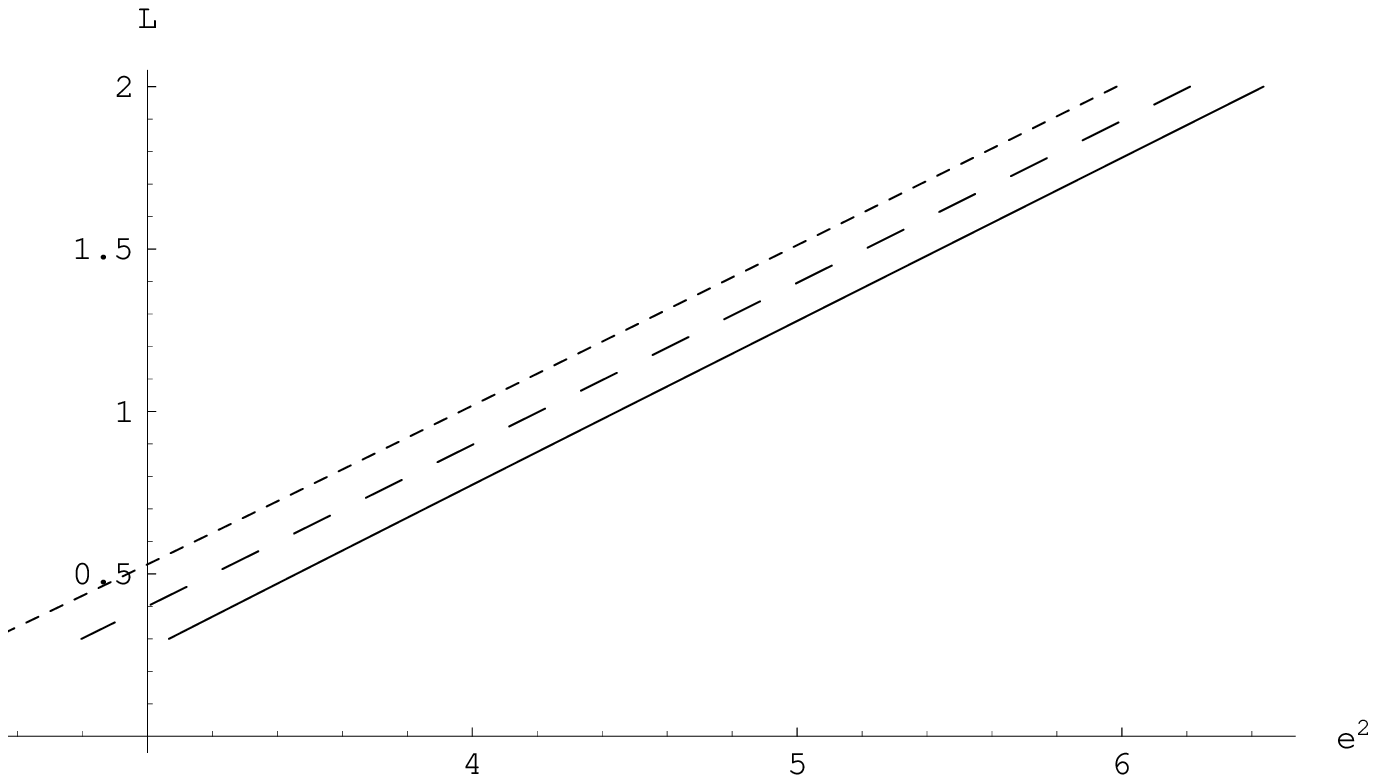}
\caption{The left dashed curve corresponds to the color Coulomb constant $k=0.4$, the middle dashed curve is for $k=0.2$ while the solid one is for $k=0$. \label{18}}
\end{figure}

\begin{figure}[ht]
\includegraphics[width=\columnwidth,angle=0]{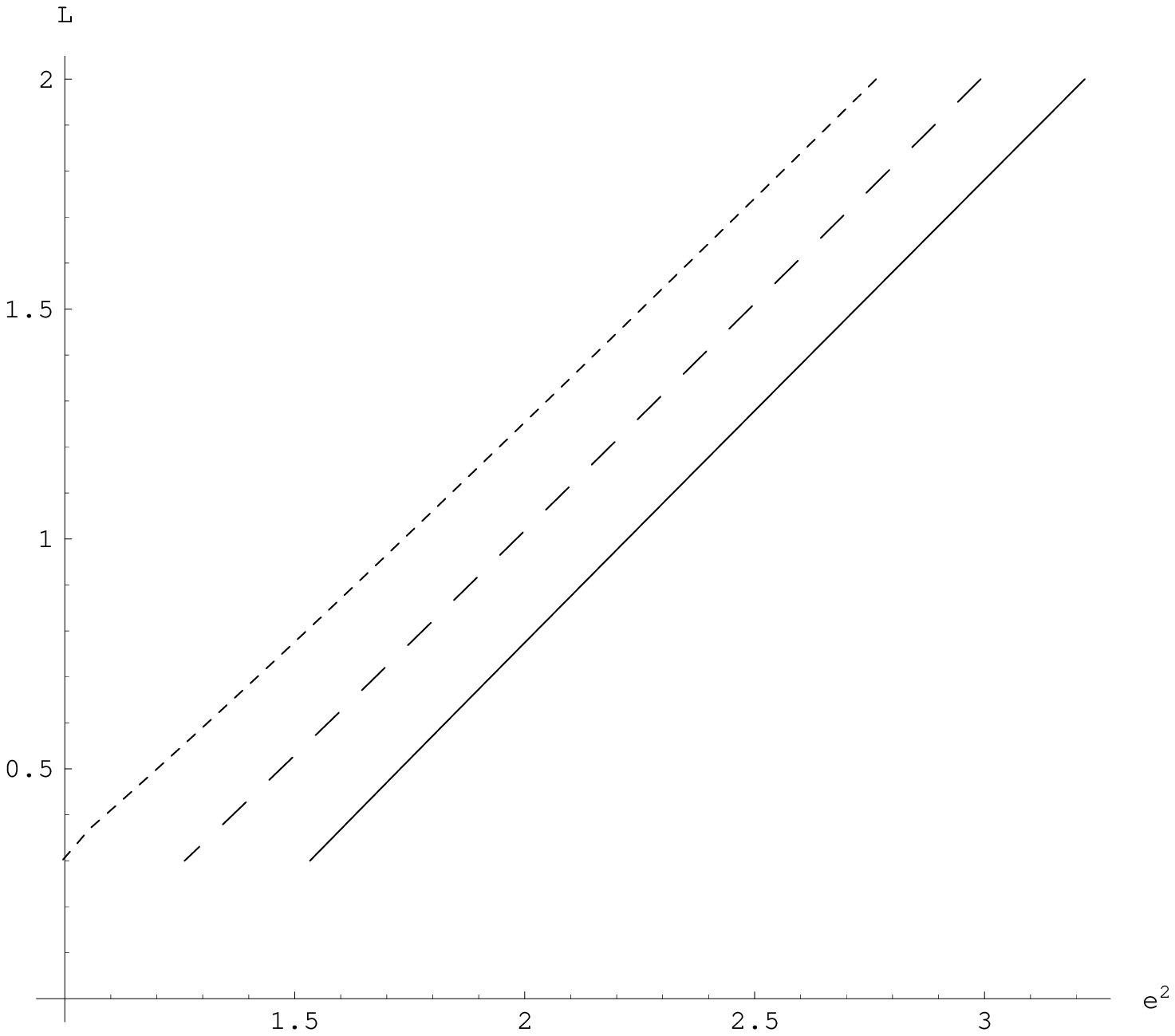}
\caption{The solid curve represents color Coulomb constant $k=0$, the left dashed curve corresponds to the $k=0.4$ while the middle dashed one is for $k=0.2$. \label{20}}
\end{figure}

Now considering the effect of color Coulomb potential and assuming that $k=0$, $k=0.2$ and $k=0.4$ separately, then we obtain the curves modified by k values in FIG.\ref{18}. Again we find that the slopes of the curve for LL mesons are half of those for HL mesons shown in FIG.\ref{20}. 

\section{Summary and Conclusion}

We have studied the relativistic flux tube model and its applications to the meson spectroscopies.  Under two limit we also solve the relativistic flux tube equations for LL mesons. We have presented the exact numerical solution for light-light mesons and compared the numerical solution with an analytical approximation solution in the deep radial limit. The comparison of solutions for LL mesons with those for HL mesons is also discussed.

We conclude that the RFT model reproduces the straight line Regge trajectories and exibit the correct string slopes for both LL and HL mesons. They appear very naturally in the RFT approach. The RFT model correctly account for the rotation of the QCD string. From the comparison of exact numerical solution with the analytic approximation solution for light-light mesons, we find that they are much more in agreement with each other for higher excited states since the deep radial limit is better satisfied. The relativistic flux tube model describes the universal relations for LL and HL meson spectroscopies correctly.

\section*{Acknowledgment}

The author would like to thank M. Olsson and C. Goebel for their helpful discussions and collaborations. This work was supported in part by the U.S. Department of Energy under Contract No.~DE-FG02-95ER40896.

\section*{Appendix    Derivation of the analytical approximation solution under deep radial limit}

We now derive an analytic approximation solution of RFT model for LL mesons under the deep radial limit $L\ll E^2$ and $k=0$. To this purpose, we define 

\begin{equation} 
\beta = \frac {2L}{e^2}
\end{equation} 

Since $\beta \ll 1$ in the deep radial limit, m and h may be represented with $\beta$ as 

\begin{eqnarray} 
m =\frac{1- \beta}{1+ \beta} \approx 1-2 \beta
 \label{eq.18}\\ 
h \approx 1- \beta
\label{eq.19} 
\end{eqnarray}

Plugging into Eq.(\ref{jjj}) we obtain

\begin{equation} 
 J = \int^1_0 dy \frac {\displaystyle\sqrt{(1-y^2)(1-y^2+2 \beta y^2)}}{\displaystyle 1-y^2+\beta y^2} 
\end{equation}

Assuming that,
\begin{equation} 
1-y^2 = \frac{1}{1+x}
\end{equation}

then $J$ may be represented with the new parameters $x$ and $\beta$ as

\begin{equation} 
 J =\frac 12 \int^\infty_0 dx \frac{\displaystyle \sqrt{\frac{\displaystyle (1+2 \beta x)}{\displaystyle x (1+x)^3}}}{\displaystyle 1+ \beta x} 
\end{equation}

Next expand $J$ with taylor expansion and drop terms of second order in $\beta$ or higher, so 

\begin{equation} 
 J= J(0)+ \frac {\partial J}{\partial \beta} \beta + \bigcirc (\beta^2)
\end{equation}
 
For $\beta=0$,

\begin{equation} 
 J(0) = \frac 12 \int^\infty_0 dx \sqrt{\frac{\displaystyle 1}{\displaystyle x(1+x)^3}} = 1
\end{equation}

and 

\begin{eqnarray}
\left.\frac {\displaystyle \partial J}{\displaystyle \partial \beta}\right|_{\beta=0} & = & \frac 12 \int^\infty_0 dx \sqrt{\frac{\displaystyle 1}{\displaystyle x(1+x)^3}} \frac{\displaystyle \partial (\frac {\displaystyle \sqrt{1+2 \beta x}}{\displaystyle 1+ \beta x})}{\displaystyle \partial \beta},  \nonumber  \\  
& = & \frac 12 \int^\infty_0 dz\frac{\displaystyle 1}{\displaystyle \sqrt{(1+2 z)} (1+z)^2} = \frac 12-\frac {\pi}{4}
\label{eq:aa}
\end{eqnarray}

where we let $z=\beta x$. Therefore when  $\beta \ll 1$, we find to the first order in $ \beta$

\begin{eqnarray}
I & = & \frac{\displaystyle  (e^2-2L) \sqrt{e^2+2L}}{\displaystyle e} J= e^2 (1-\beta)(1+\beta)^\frac 12 [1+(\frac 12 -\frac {\pi}{4})\beta\,] \nonumber  \\  
& = & e^2 (1-\frac {\pi}{4}\beta)
\label{eq:ab}
\end{eqnarray}

After substituting $e^2$ with $\frac{E^2}{4a}$, we obtain an analytical approximation of the heavy-light semi-classical time component vector potential integral,

\begin{equation} 
I =\frac {E^2}{2a} (1-\frac {\pi}{4}\beta)
\end{equation} 

Applying the semi-classical quantization condition for a spherically symmetric system  and replacing the classical angular momentum with Langer correction to take into account the centrifugal singularity, we immediately obtain the spectroscopic relation

\begin{equation} 
L + 2 n+ \frac 32 =\frac{E^2}{2\pi a} 
\end{equation}

which is the Eq.(\ref{dd}) and an excellent approximation to the numerical solution of RFT model for a LL meson presented in section \ref{ana}.

\end{document}